\documentclass{IOS-Book-Article}

\usepackage{mathptmx}
\usepackage{graphicx}
\usepackage{hyperref}
\usepackage[dvipsnames]{xcolor}

\def\hb{\hbox to 10.7 cm{}}

\newcommand{\pkg}[1]{{\normalfont\fontseries{b}\selectfont #1}}
\let\proglang=\textsf

\begin{document}

\pagestyle{plain}
\def\thepage{}

\begin{frontmatter}              

\title{The Literary Theme Ontology for Media Annotation and Information Retrieval}

\markboth{}{August 2019\hb}

\author[A]{\fnms{Paul} \snm{Sheridan}%
\thanks{Corresponding Author: Chief Science Officer, Tupac Bio, Inc., 717 Market St.,
San Francisco, CA, 94103; E-mail: paul.sheridan.stats@gmail.com.}},
\author[B]{\fnms{Mikael} \snm{Onsj{\"o}}}
and
\author[C]{\fnms{Janna} \snm{Hastings}}

\runningauthor{P. Sheridan et al.}
\address[A]{Tupac Bio, Inc., USA}
\address[B]{Independent Researcher, UK}
\address[C]{Department of Clinical, Educational and Health Psychology, University College London, UK}

\begin{abstract}
Literary theme identification and interpretation is a focal point of literary studies scholarship. Classical forms of literary scholarship, such as close reading, have flourished with scarcely any need for commonly defined literary themes. However, the rise in popularity of collaborative and algorithmic analyses of literary themes in works of fiction, together with a requirement for computational searching and indexing facilities for large corpora, creates the need for a collection of shared literary themes to ensure common terminology and definitions. To address this need, we here introduce a first draft of the Literary Theme Ontology. Inspired by a traditional framing from literary theory, the ontology comprises literary themes drawn from the authors own analyses, reference books, and online sources. The ontology is available at \url{https://github.com/theme-ontology/theming} under a Creative Commons Attribution 4.0 International license (CC BY 4.0).
\end{abstract}

\end{frontmatter}
\markboth{August 2019\hb}{August 2019\hb}

\section{Introduction}
Literary themes are the notable topics that structure and give focus to works of fiction, by relating art to life around such themes as ``romantic love'', ``coping with death'', and ``the lust for power'', or to the science fictional and fantastical in themes ranging from ``astronomical-scale engineering'' projects to ``the occult''. They are a natural organising hub for classification and retrieval of works of fiction. 

Despite many informal textual collections of literary themes, no comprehensive \textit{ontology} has yet been developed to underpin thematic analysis in digital literary studies. Ontologies are computationally formalised representations of entities from a given domain. By capturing domain knowledge in a standardised computable form, they enable sophisticated applications such as clustering and intelligent information retrieval. Based on best practices such as the use of the Basic Formal Ontology (BFO)~\cite{Smith2005} as upper level and the Web Ontology Language (OWL2)~\cite{Hitzler2009} as encoding formalism, we here introduce, and present a first draft of, the Literary Theme Ontology (LTO). 

The LTO has been developed to be a general-purpose literary theme ontology able to serve multiple applications. However, in this initial presentation we focus in the main on science fiction themes, largely derived from a case study of the analysis of the \textit{Star Trek} television series episodes, and the application scenarios of classification and information retrieval. The LTO is available for download at \url{https://github.com/theme-ontology/theming} under a Creative Commons Attribution 4.0 International license (CC BY 4.0). For user-friendly access to the current developmental version of the ontology along with a searchable database of annotated stories visit the related Theme Ontology project website (\url{https://themeontology.org})~\cite{TO2019}.

\section{Materials and Methods}

\subsection{Literary Themes}

It is customary among literary theorists to distinguish between subject and literary theme~\cite{Griffith2010, Murfin2009}. According to this view, a subject of a literary work is a topic that is explored within that work. A literary theme, on the other hand, is an opinion that a work conveys about one of its subjects. Kelley Griffith illustrates this distinction by noting that ``love'' is a subject of Shakespeare's \textit{Sonnet 116} and ``love remains constant even when assaulted by tempestuous events or by time'' a literary theme~\cite{Griffith2010}. To take another example, in Jules Verne's \textit{Twenty Thousand Leagues Under the Sea} (1870), ``the obsessive pursuit of retribution inevitably leads to wreck and ruin'' is arguably a literary theme associated with the subject ``vengeance''.

For the purposes of the LTO, a literary theme of a work of fiction is considered to be a topic treated by that work or an opinion conveyed by the work about a topic treated therein. In other words, we use the term ``literary theme'' to encompass both the subjects and literary themes of classical literary theory. Several practical considerations motivate this choice. In the first place, the above-drawn distinction between literary theme and subject is far from universally observed. To single out one example, C.E. Preston and J.A. Cuddon write of ``jealousy'' (a subject) being a literary theme of Shakespeare's \textit{Othello}~\cite{Preston1998}. Second, in practice, subjects are more susceptible to objective characterization, than the comparatively subjective classical literary themes associated with them. That said, we do wish to accommodate the identification of classical literary themes whenever feasible. Third, subjects sometimes evoke an implicit theme to begin with. For example, the subject ``Martian extraterrestrial'' evokes the associated theme ``the planet Mars is home to sentient beings'' without too much stretching of the imagination. From the perspective of classification and information retrieval, little is gained by distinguishing between topics and the opinions a work coveys about them in such cases. In the authors' judgement, the more profitable classical literary themes to include in the ontology amount to aphorisms, such as ``appearances can be deceiving'', ``a danger shared can bring people together'', and ``be careful of what you wish for'' to name but a few.

\subsection{Literary Theme Acquisition and Curation}

The LTO terms were primarily collected by watching science fiction video productions and recording themes as the authors understood them. We refer readers to~\cite{Onsjo2019, Sheridan2019} for a discussion of the methodology we used to populate the ontology with themes. Table~\ref{TAB:lto-documents} catalogs the lion's share of works covered in the ontology to date. The heavy focus on \textit{Star Trek} is explained by its being a long-running, culturally significant franchise with a serviceable balance of themes relating to the human condition, society, and futuristic technologies. Although it will also be admitted that the authors are particular to the show. In order to cover the pre-\textit{Star Trek} era, we watched and recorded themes for roughly $500$ classic science fiction films, dating from $1895$ to $1973$, drawn from Wikipedia lists of science fiction films by decade~\cite{wikipedia2019b}. Note that \textit{Star Trek} first aired in $1966$. We performed the same analysis for about $100$ selected science fiction films from $1974$ through to the present day. On top of that we watched and recorded themes for the \textit{Black Mirror} television series and the animated series \textit{Futurama}, in an effort to capture some of the more modern-day science fiction themes. Lastly, we are at present in the process of synthesizing and semantically structuring existing compiled lists of heterogeneous themes from \textit{The Encyclopedia of Science Fiction}~\cite{SFE2019}, \textit{Dictionary of Literary Themes and Motifs}~\cite{Seigneuret1988}, \textit{Encyclopedia of Themes in Literature}~\cite{McClintonTemple2010}, \textit{Encyclopedia of Film Themes, Settings and Series}~\cite{Armstrong2001}, and scattered online sources~\cite{wikipedia2019a, LD2019}.

\begin{table}[!ht]
\caption{Science fiction television series and films used to populate LTO with terms.}
{\label{TAB:lto-documents}}
\centering
\begin{tabular}{lcr}
\hline
Story Collection & Time Period & Story Count \\
\hline
 Classic science fiction films & 1895\thinspace --\thinspace 1973 & 491 \\
 Selected science fiction films & 1974\thinspace --\thinspace 2018 & 105 \\
 Star Trek: The Original Series & 1966\thinspace --\thinspace 1969 & 79 \\
 Star Trek: The Animated Series & 1973\thinspace --\thinspace 1974 & 22 \\
 Star Trek: The Next Generation & 1987\thinspace --\thinspace 1994 & 178 \\
 Star Trek: Deep Space Nine & 1993\thinspace --\thinspace 1999 & 177 \\
 Babylon 5 & 1994\thinspace --\thinspace 1998 & 110 \\
 Star Trek: Voyager & 1995\thinspace --\thinspace 2001 & 172 \\
 Futurama & 1999\thinspace --\thinspace 2013 & 129 \\
 Star Trek: Enterprise & 2001\thinspace --\thinspace 2005 & 99 \\
 Black Mirror & 2011\thinspace --\thinspace 2019 & 23 \\
\hline
\end{tabular}
\end{table}

The scope of the LTO is the representation of operationally verifiable themes of some significance that can be expected to arise in multiple stories. The term \textit{story} is used in the present work to refer to any work of fiction, be it a work of written literature, a film, a television drama, comic, video game plot, and so on. 

Theme identification is a subjective enterprise. The LTO has been designed with an understanding that some important themes will fall necessarily outside of its scope on account of their ambiguous or controversial nature. Indeed, people may reasonably disagree because people with different perspectives are liable understand stories differently. Instead, the LTO's focus is squarely on themes which are amenable to being defined in precise terms. Take the timeless theme ``the desire for vengeance'' (alias ``vengeance'') as an example. It is defined as ``A character seeks retribution over a perceived injury or wrong.''. The classic science fiction theme ``flying car'' constitutes another example. It is defined as ``There is a hypothetical air vehicle that looks like a car.''. Both themes are relatively straightforward to identify in stories. We minimize this subjective element of story theme identification by insisting on maximally unambiguous definitions.

\subsection{Literary Theme Ontology Construction}

In compiling the LTO  from the above mentioned collection of themes, we followed BFO best practices to ensure interoperability with existing and future humanities related ontologies. The ontology was constructed using Prot\'{e}g\'{e} version 5.5.0~\cite{Musen2015} in the Web Ontology Language (OWL2)~\cite{Hitzler2009}. We captured themes as classes in OWL, related using \textit{subClassOf}, and annotated with the annotation property \textit{label} from the RDF Schema. In addition, we used the Open Biological and Biomedical Ontologies (OBO)~\cite{Smith2007} metadata standard annotation properties \textit{definition}, \textit{alternative term}, and \textit{reference}. The Fast Classification of Terminologies FaCT++~\cite{Tsarkov2006} logic reasoner shows that the LTO is formally consistent and contains no cycles.

\subsection{Release and Quality Control}

Current and archival releases of LTO are made available for download at the project's GitHub repository (\url{https://github.com/theme-ontology/theming}) under a Creative Commons Attribution 4.0 International license (CC BY 4.0). In the future, we aim to recruit a community of domain experts (i.e. digital literary scholars) to quality check all themes in the ontology.

\subsection{Theme Requests}

Technical support and new theme requests are presently handled at the LTO GitHub repository (\url{https://github.com/theme-ontology/theming}).

\section{Results}

\subsection{LTO Overview}

\begin{figure}[!ht]
\centering
\includegraphics[width=\columnwidth]{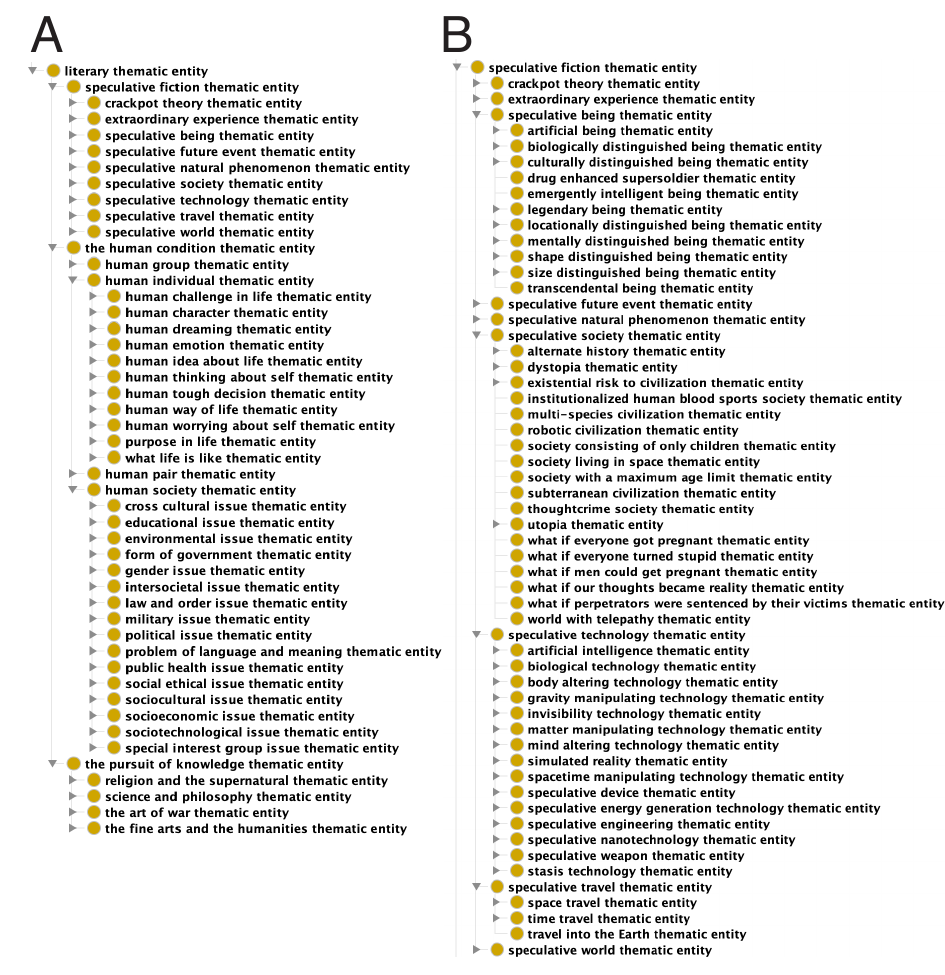}
\caption{A) LTO bird's eye view. B) LTO with focus on speculative fiction themes.\label{FIG:lto-overview}}
\end{figure}

The LTO version 0.3.1 contains $2,875$ classes (i.e. literary themes). The upper-level organization is loosely based around a literary theme classification scheme proposed by William Henry Hudson in 1913~\cite{Hudson1913}. He identifies five sweeping categories, which we quote:
\begin{enumerate}
    \item the personal experiences of the individual as individual\thinspace ---\thinspace the things which make up the sum-total of his private life, outer and inner;
    \item the experiences of man as man\ldots which transcend the limits of the personal lot, and belong to the race as a whole; 
    \item the relations of the individual with his fellows, or the entire social world, with all its activities and problems;
    \item the external world of nature, and our relations with this; and
    \item man's own efforts to create and express under various forms of literature and art.
\end{enumerate}
Figure~\ref{FIG:lto-overview}A shows an overview of the LTO upper level themes. The root class ``literary thematic entity'' has three children: ``the human condition thematic entity'', ``knowledge and belief thematic entity'', and  ``speculative fiction thematic entity''. As these class names suggest, theme names in the LTO follow the convention ``such and such a theme thematic entity''. However, the ``thematic entity'' phrasal appendage is dropped in the exposition below for brevity's sake. Each term is accompanied by a definition, and a reference when possible.

The ``the human condition'' theme covers the first and third of Hudson's five categories. In keeping with Hudson, the LTO covers human experience at expanding levels of interaction. First, ``human individual'' covers themes about characteristic key events, and situations which compose the essentials of human experience, such as birth and death, growth, emotionality, aspiration, conflict, and life decisions. Second ``human pair'' themes examine the respective interactions and attitudes of two people vis-\`{a}-vis each other. This theme is specialized in such descendants as ``mother and daughter'', ``husband and wife'', and ``friendship'' (i.e. ``friend and friend''). Third, ``human group'' covers themes pertaining to small groups of people in which everyone effectively knows everyone else (e.g., family, friends, and the workplace). And last, ``human society'' covers themes about how groups of people in the same geographical territory subject to the same political authority and cultural expectations organize themselves and interact with other such entities. In the future, these levels of human interaction can be extended according to the ``levels of analysis'' approach used in sociology~\cite{Babbie2004}.

The upper level class ``the pursuit of knowledge'' functions as a generalization of Hudson's three remaining theme categories. In practice, this is a home for themes pertaining to the expression of a view about how the world operates, and how humans fit in relation to it. Put another way, these are themes about scientific, religious, philosophical, artistic, and humanist views on the nature of reality.

The last upper level LTO class, which is unaccounted for in Hudson's classification scheme, is ``speculative fiction thematic entity''. Figure~\ref{FIG:lto-overview}B shows an overview. The class covers subject matter either falling outside of reality as it is presently understood or to human experiences of such an extraordinary nature that no reasonable person would expect to experience them in the course of their lifetime. Themes of these sorts are prevalent in, but are not exclusive to, the genres of science fiction, fantasy, adventure fiction, and horror. The theme ``Dyson sphere'', defined as ``There is a truly enormous, hollow sphere constructed around a star to capture its entire energy output.'', constitutes a typical example of a theme-worthy entity that does not presently exist in our universe so far as anybody knows. Meanwhile, the theme ``what if I were stranded on a deserted island'' serves as a stock example of a situation that historical figures, notably Alexander Selkirk~\cite{Howell1829}, actually experienced, but is so extraordinary that one would be utterly astonished to find oneself in the same unenviable predicament.

\subsection{Alignment with BFO}

To ensure interoperability with other ontologies, including other literature-related ontologies that might be developed in the future, the LTO was designed to fit within the BFO class hierarchy~\cite{Arp2015}. While BFO has mainly been developed to provide an upper level ontology for \textit{scientific} domains, and has not specifically been developed to categorise fictional entities or themes, we contend that it is nevertheless not incompatible with our aims. BFO provides a broad metaphysical framework that gives structure and definitions for many of the general distinctions between types of entity that cut across domains and subjects, based on different ontological modes of existence. For example, BFO distinguishes between entities that exist as stand-alone entities and those that are dependent on other entities for their existence. An example of the former is an apple, while the red colour of an apple is an example of the latter, since the colour cannot exist without the apple of which it is the colour. 

We take as our precedent for considering literary themes as BFO entities, the \textit{information content entities} that are the subject matter of the Information Artifact Ontology (IAO)~\cite{Ceusters2012}. Information content entities are \textit{dependent} entities in BFO, in that they cannot exist without some bearer. The bearer of an information content entity is its \textit{concretization}. For example, the content may be concretized by printing it on paper (e.g. a book) or it may be concretized in a file on a computer (e.g. a PDF file). The information content entity is not identical with its concretizations, but is rather multiply realized by them (in BFO terms, it is \textit{generically dependent}) insofar as the same (for example) book \textit{content} can be reproduced in multiple copies. 

Information content entities are intended to describe the realm of data and scientific research reports, rather than fiction. They are defined by a relation of \textit{is about}, or representation: information content entities are \textit{about} some entity in the real world. A single data item in a database may be about an individual entity in the world, but more typically information content entities are about complex aggregates of entities and interrelations, which may be called \textit{portions of reality}. 

\begin{figure}[!ht]
\centering
\includegraphics[width=\columnwidth]{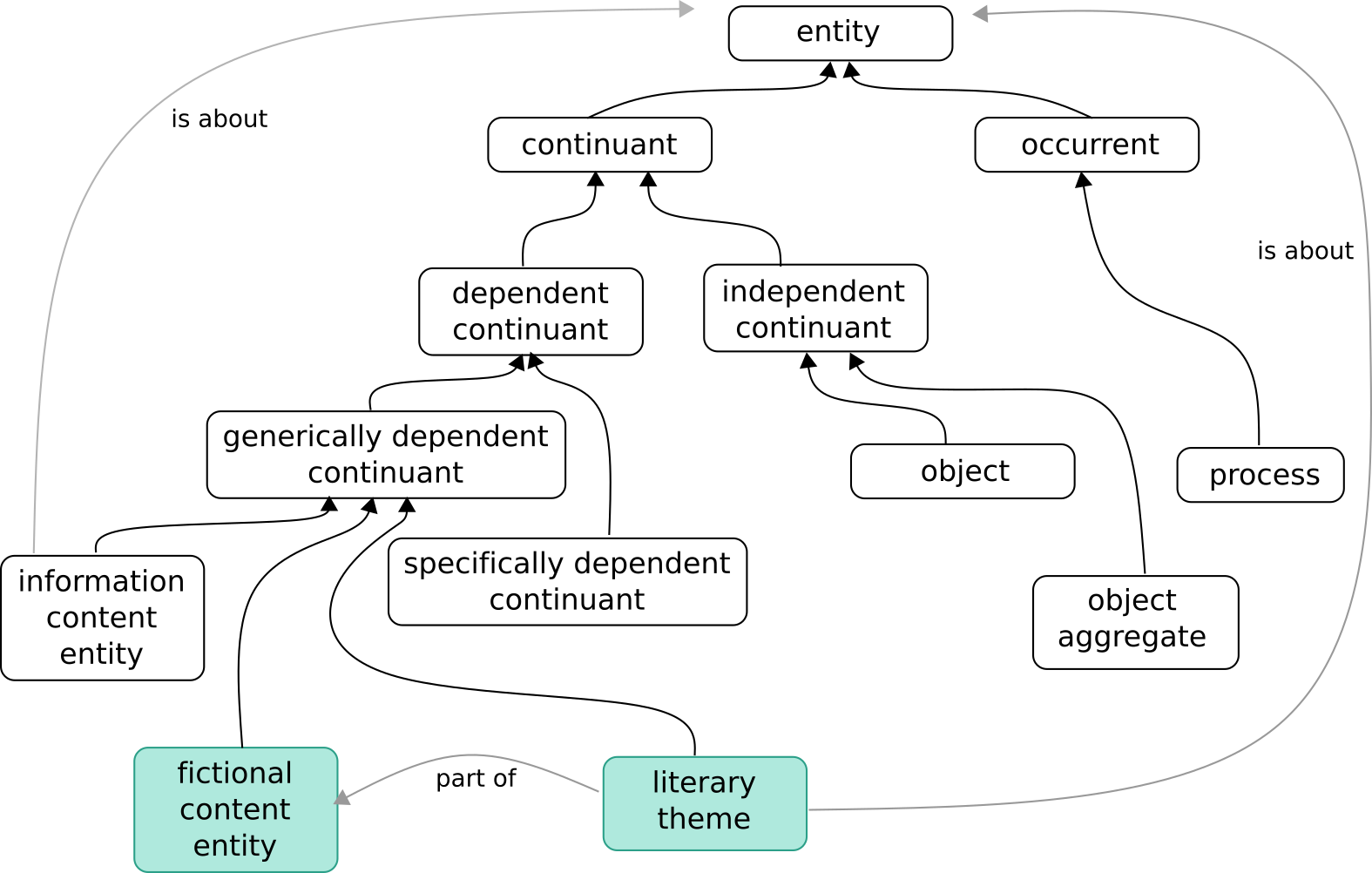}
\caption{Literature themes in the context of the BFO class hierarchy. Black boxes indicate BFO classes, while coloured boxes indicate proposed classes relevant to literature and literary themes. Black arrows represent hierarchical (is a) relations, while labelled grey arrows represent other types of relation.\label{FIG:lto-bfo}}
\end{figure}

Works of literature are not exactly information content entities, since they are typically not intended to carry information about portions of reality in the same way. (Although, of course, in some genres, such as historical fiction, the boundaries may be blurred.) Nevertheless, it should be clear that there are many similarities: works of fiction are also \textit{content} entities, generically dependent on their bearers in which they are concretized, able to be copied and reproduced in different concrete forms. However, they are not \textit{information} content entities, but rather \textit{fictional} content entities. The distinction amounts to a difference in the nature of the type of representation involved: while information content entities are \textit{about} portions of reality quite straightforwardly, works of literature only \textit{appear to be} about portions of reality. Fictional entities have a peculiarly dual ontological aspect: an apparent nature (within the fictional work) and a real nature (as a fictional work), as discussed at length, for example, in \cite{Ingarden1973}. Thus, we categorise works of fiction as \textit{dependent continuants} in BFO, alongside information content entities. Figure~\ref{FIG:lto-bfo} illustrates this proposal.

As discussed above, we define literary themes pragmatically as subjects, or opinions or judgements about subjects, expressed in works of literature. In this broad sense, literary themes can be understood as \textit{part of} the broader literary content entity, and are often \textit{about} (in the IAO sense) the world in which the work of literature is situated. For example, a literary theme such as ``war'', or ``romantic love'', is straightforwardly \textit{about} the class war or romantic love as understood as a universal in the world. However, not all literary themes are about entities in the world and thus this relation is not an essential one.  

\subsection{Comparisons with Related Knowledge Curation Systems}

The LTO constitutes, to the authors' knowledge, the first comprehensive ontology of literary themes. Traditional scholarship in literary studies occasioned numerous reference works containing lists of themes~\cite{Seigneuret1988, McClintonTemple2010, Armstrong2001}. Themes lists also abound on the web. The Encyclopedia of Science Fiction website~\cite{SFE2019}, which features an invaluable list of science fiction themes, is one example of immediate relevance to the LTO. Moreover, there is nowadays a growing number of wikis focused on the curation of terms that overlap in scope with the LTO, including, TV Tropes~\cite{tvtropes2019}, the Alien Species Wiki~\cite{alienspecies2019}, and Non-alien Creatures Wiki~\cite{nonaliencreatures2019}. The terms on wikis are typically organized into categories, rather than fully fleshed out hierarchies.

Ontologies have been and continue to be used to model information in the humanities. Relevant to literary and film studies is the \textit{Noctua literaria} ontology, which is an ontology of literary characters~\cite{zoellner2009}. Bartalesi and Meghini~\cite{bartalesi2017} developed a semantic network-style ontology in RDF for representing knowledge contained in literary texts with emphasis on the works of Dante Alighieri. The ontology features 60 thematic areas (e.g. astronomy, occultism, and theology). More recently, Fabio Ciotti has advanced a more general ontological framework of narrative that includes characters as a special case~\cite{Ciotti2016}. OntoMedia is a long-established ontology that includes a hierarchical classification scheme quite relevant to setting~\cite{Jewell2005}. And most recently, Damiano~et~al. have published The Ontology of Drama~\cite{Damiano2019}, dubbed Drammar, with a view toward formally encoding the domain of dramatic media. Lastly, Khan~et~al.~\cite{Khan2016} reorganized an existing taxonomy of literary themes and motifs to fit within the framework of the DOLCE-lite upper level ontology. The ontology is implemented in OWL. The underlying literary themes and motifs were compiled by experts for the purpose of annotating Ancient Greek and Latin texts. There is overlap with the LTO in such timeless themes as ``love'', ``death'', and ``war''. But there are interesting differences also. For example, a cursory inspection of their ontology reveals the LTO to be conspicuously lacking in the different types of physical metamorphoses. Khan~et~al. record ``metamorphosis in to animals'', ``metamorphosis in to flowers'', ``metamorphosis in to fruit'', and ``metamorphosis in to trees''. By contrast, the LTO contains the ``what if I underwent a physical metamorphosis'' themes: ``what if I grew in size'', ``what if I shrank in size'', and ``what if I were gradually turning into stone''. In the future, the LTO stands to be improved by incorporating a variety of these themes.

\subsection{LTO Applications}

The LTO covers a broad range of themes that can be expected to recur in works of literature, films, television and online programs, and other media. Consequently, we envision a number of knowledge curation and information retrieval applications.

Concerning knowledge curation, LTO functions as an ontology-based vocabulary that can be used to tag stories with themes in a manner that exploits its hierarchical organization. Organizing tags into different levels of abstraction helps facilitate tagging by compartmentalizing ambiguity. Consider the subject of aliens, to take one example. In the classic sci-fi film \textit{The Day the Earth Stood Still} (1951), a nondescript, humanoid alien comes to Earth to deliver an important message to humanity concerning the dangers of atomic weapons. The theme ``extraterrestrial being'' (a child of ``locationally distinguished being'') applies, since there is nothing particularly noteworthy about him beyond the fact that he comes from outer space. However, it is possible to use a more specialized theme when appropriate. For example, ``Venusian extraterrestrial'' in stories featuring beings from the planet Venus, or something more exotic, like ``mysterious maker alien race'' should the occasion demand it. Both themes descend from ``extraterrestrial being''. The lesser known film \textit{Last Woman on Earth} (1960) presents another illustrative case study. In the film, all animal life on Earth, save for three people who were diving under the sea, appeared to have perished after the oxygen suddenly disappeared from the atmosphere. What caused the de-oxygenation event was never explained. But plants soon thereafter began to replenish the atmosphere with oxygen so that the three lucky survivors could breathe again. We deemed it appropriate to tag the story with the high level theme ``existential risk to civilization'', since the process by which Earth's oxygen disappeared is left open to speculation. A final general example is the ``international issue'' of the use of ``weapons of mass destruction'' in war. LTO presently features ``biological weapons'', ``chemical weapons'', and ``nuclear weapons'' as ``weapons of mass destruction'' themes. These can be used to tag stories at different levels of granularity as required.

There is a host of information retrieval applications. A statistical test making use of an earlier version of the LTO was developed to identify over-represented themes in stories of interest relative to a background list of stories~\cite{Onsjo2019}. An ontology-enhanced fiction content recommender system based on another earlier version of the LTO has also been proposed~\cite{Sheridan2019}. Future applications include LTO ontology-enhanced document clustering~\cite{Fahad2017} methods for clustering stories together into subtypes based on themes they share in common, and time-course analysis methods to identify themes whose usages differ significantly between two group of stories over time (e.g. author gender or nationality). Yet another potential application is to use LTO themes as a gold standard for document classification tasks in conjunction with inter-rater concordance to assess the classification quality. The larger our corpus of thematically annotated stories grows over time, the more powerful these sorts of statistical methods will become for systematically exploring story data, testing existing hypotheses, and generating new hypotheses with the potential to offer novel insights into the real world.

Lastly, automated methods have been developed for extracting themes from texts. Topic modeling techniques~\cite{Blei2012} as implemented in such software packages as MALLET~\cite{McCallum2002}, the Python module gensim~\cite{Rehurek2010}, topicmodels~\cite{Grun2011}, and the \proglang{R} package \pkg{tm}~\cite{Feinerer2008} have been used successfully to identify literary themes in large textual copora~\cite{Jockers2013a, Jockers2013b, Goldstone2014, Boyd-Graber2017}. An  interesting challenge awaits in adapt the current methods for automatic topic labeling~\cite{Lau2011, Basave2014, Bhatia2017} to the problem of mapping identified topics to themes in the LTO.

\section{Discussion}

We have presented a first draft of an ontology of literary themes. The LTO, as we call it, is a natural extension of traditional literary theme lists found in reference books~\cite{Seigneuret1988, McClintonTemple2010, Armstrong2001} and newly emerging wiki-style, collaborative trope curation websites~\cite{tvtropes2019, alienspecies2019, nonaliencreatures2019}. The LTO is the first comprehensive ontology of literary themes. It is designed to be BFO compliant, implemented in the OWL2 ontology language, and features upper classes based on a traditional literary classification scheme advanced by William Henry Hudson~\cite{Hudson1913}. The LTO version 0.3.1 is comprised of $2,875$ literary themes that can be expected to cover stories across a wide range of different genres and media types. 

The version of the LTO presented in this paper uses OWL2 as a formalism to express a taxonomy of literary themes. An important reason for developing the LTO as an ontology, instead of just as a taxonomy or vocabulary, is to capitalize on the established role ontologies play in the development of community-wide shared standards. While taxonomies can and often do serve as community standards, ontologies are explicitly designed for that purpose and there are communities, such as the OBO Foundry~\cite{Smith2007}, with a wealth of expertise on best practices for the development of ontologies to serve as standards. On top of that we plan to extend the LTO beyond a taxonomy via the incorporation of additional relations. For example, ``what if I shrank in size'' (an ``extraordinary experience'') and ``miniturization technology'' (a ``speculative technology'') could both be linked to a non-thematic appropriately defined processual entity ``shrinking'' by the \textit{is-about} relation, to facilitate querying the ontology for all themes related to things becoming smaller in size. We furthermore aspire to add full logical definitions for themes to the LTO whenever possible.

It is our aim to improve upon the LTO to the point where it can function as a serviceable community standard for researchers involved in the digital annotation of story corpora. A most urgent future work, therefore, is for the LTO to be critiqued by digital literary scholars, and revised accordingly. On the subject of annotation, we compiled a modest collection of annotated science fiction media as a proof of concept. Future work will begin with the drafting of a workable protocol for the collaborative tagging of stories with LTO themes. Aligning the LTO with other media ontologies is another future work of some importance. Already existing ontologies, such as OntoMedia~\cite{Jewell2005}, are designed with a view toward annotating media objects according to characters, settings, and other qualities.

LTO development is currently managed at the project's GitHub repository (\url{https://github.com/theme-ontology/theming}). We encourage anyone interested in contributing to the development of LTO to submit an outside collaborator request to the Theme Ontology GitHub organization. Project collaborator contributions (e.g. proposals for new themes, suggested updates to existing themes, annotated stories) will be subject to approval by a repository manager according to the standard GitHub pull request workflow. In the long term, we aim to transform the Theme Ontology website (\url{https://themeontology.org}) into a thriving online community platform by following the non-profit model of Wikipedia. At the start we envision that contributors will be required to register with either an academic email address or by referral from an existing member. Wiki policies and guidelines remain to be drafted and a collaborative story annotation protocol fully formulated. But community members will ultimately be able to annotate stories with LTO themes, propose new themes, update or delete old ones, and improve on the ontology structure\thinspace --\thinspace all based on a wiki-style consensus model. The more the Theme Ontology database grows with LTO thematically annotated stories, the more the sorts of information retrieval methods outlined above can be harnessed to test hypotheses about human culture and society.

To close on a more general note, the biological and medical sciences enjoy stable platforms for ontology management~\cite{Smith2007, Musen2011}. However, no such system enjoys wide adoption in the humanities, to the authors knowledge. There is a pressing need for the creation of a platform with the goal of organizing interoperable reference ontologies in the digital literary studies domain.

\bibliographystyle{ios1}
\bibliography{references}

\end{document}